\documentclass{article}

\usepackage[dvipsnames,svgnames]{xcolor}

\usepackage[english]{babel}
\usepackage{microtype}
\usepackage{tikz}
\usetikzlibrary{calc}
\usepackage{pgfplots}
\pgfplotsset{compat=newest}
\usepackage{pgfplotstable}
\usepackage{subcaption}  
\usepackage{caption}
\usepackage{amsfonts}
\usepackage{amsmath}
\usepackage{algorithm}
\usepackage{algpseudocode}
\usepackage{authblk}

\usetikzlibrary{arrows.meta,positioning,shapes.geometric}

\tikzset{
  fc/line/.style       = {-Latex, thick},
  fc/block/.style      = {rectangle, rounded corners=2pt, draw, align=center,
                          minimum width=3.5cm, minimum height=1.1cm},
  fc/decision/.style   = {diamond, aspect=2, draw, align=center, inner sep=1.5pt},
  fc/term/.style       = {ellipse, draw, align=center, minimum width=2.8cm, minimum height=1.1cm}
}

\usepackage{comment} 
\providecommand{\keywords}[1]
{
  \small	
  \textbf{\textit{Keywords---}} #1
}
\title{Exceeding the Numerical and Performance Characteristics of IEEE-754 SGEMM with BFloat16 Tensor Cores on GPUs for Scientific Computing}

\author[1] {Harun Bayraktar, hbayraktar@nvidia.com}
\author[1] {Cole Brower, cbrower@nvidia.com}
\author[1] {John Gunnels, jgunnels@nvidia.com}
\author[1] {Greg Henry, grhenry@nvidia.com}
\author[1] {Cherin Joseph, cjoseph@nvidia.com}
\author[1] {Jack Kosaian, jkosaian@nvidia.com}
\author[1] {Dmitry Lyakh, dlyakh@nvidia.com}
\author[2] {Lukas Mosimann, lmosimann@nvidia.com}
\author[3] {Victor Podlozhnyuk, vpodlozhnyuk@nvidia.com}
\author[4] {Addison Richards, addisonr@nvidia.com}
\author[1] {Paul Springer, pspringer@nvidia.com}
\author[1] {Haicheng Wu, haichengw@nvidia.com}
\affil[1]  {NVIDIA Corporation, Santa Clara, California, USA}
\affil[2]  {NVIDIA Corporation, Zurich, Switzerland}
\affil[3]  {NVIDIA Corporation, Reading, UK}
\affil[4]  {NVIDIA Corporation, Toronto, Canada}

\begin{document}

\maketitle

\begin{abstract}
Largely due to their increased native capacity for numerical intensity and power efficiency, reduced-precision floating-point computing resources, primarily used in artificial intelligence (AI) applications, have expanded at a greater rate than their higher-precision relatives.  This has led to various efforts focused upon leveraging plentiful reduced-precision hardware to mimic higher-precision mathematical calculations.  This paper studies a specific use case, namely the use of bfloat16 (BF16) Tensor Cores found on modern GPUs in service of single precision (FP32) matrix multiply operations.  Given that BF16 and FP32 share the same dynamic range, the option to accumulate BF16 operations into FP32 accumulators (at full-speed), and additional BF16 arithmetic characteristics specific to the Blackwell GPU architecture, such as integrated scaling hardware, such emulation is highly motivated.  This paper examines the performance, efficiency, power, and numerical characteristics of FP32 matrix multiplication via BF16-based emulation and demonstrates how it exceeds numerical and performance characteristics of native FP32 for scientific applications. We also discuss a full library-ready implementation that correctly deals with denormals.
\end{abstract}




\section{Introduction}\label{sec:intro}


While scientific computing often uses double precision computation in order to retain high accuracy, it has long been known that many important domains do not require that level of precision and can avail themselves of the speed and memory savings available through the use of single precision computation and storage.  Thus, SGEMM~\cite{DonCHH88a} is a vital component in scientific computing and a great deal of time and effort has gone into optimizing its performance for many different architectures.  

With the explosive growth of AI, especially large language models (LLMs)~\cite{vaswani2023attentionneed} and generative AI~\cite{goodfellow2014generativeadversarialnetworks}, over the last decade, we have also been the beneficiaries of a good deal of expansion in the number of native data types made available on modern GPUs.  As these algorithms make extensive use of matrix multiplication, it is not surprising that reduced-precision compute capabilities, realized in hardware such as Tensor Cores, have greatly increased in terms of throughput, functionality, and energy efficiency, over the same time period.  

Given these hardware riches and the motivation to leverage them to greatest advantage for scientific computing, 
a good deal of effort has gone into research and engineering efforts for doing so.  These endeavors can be separated largely into two categories, mixed-precision approaches, usually involving iterative refinement, and emulation, wherein lower-precision units are used to imitate (emulate) the operations typically associated with higher-precision computational entities.  The former requires code intervention and, with rare exception, algorithmic ingenuity to employ, while the latter can be leveraged with little effort, via libraries, in applications.  The latter pathway is the focus of this paper. 

In this work, we seek to emulate SGEMM with a GEMM based on triplets of BF16s, and to provide the user with a familiar interface. By doing this, we enable a wide range of developers to take advantage of the superior performance, power, and efficiency, while still maintaining comparable accuracy. All the user needs to do is opt-in with an environment variable. 

While there are works that change the exponent range, we specifically focus our work with 8-bit exponent ranges of FP32. Other precisions can be used, but we focus exclusively on the same data range as FP32. 

This paper is divided into several sections. 
In Section 2, we discuss related work. 
In Section 3, we then focus on how the contributions listed below lead to algorithmic advancements. 
In Section 4, we discuss the details of our algorithm. 
Sections 5 and 6 deal with both the numerical aspects of our FP32 emulation and the application, performance, and power results reflecting where emulation yields benefits. 

This paper's contributions are:
\begin{enumerate}
\item Production-level Robust Library: Demonstrated that 9-term BF16 emulation (not reduced variants) achieves net speedups on modern hardware. Complete IEEE-754 level accuracy   including denormal   handling, lossless conversion and full FP32 exponent-range, enabling drop-in library use. 
\item Hardware Algorithm Codesign: Blackwell's tcgen05.mma instruction provides hardware-accelerated scaling during accumulation. Our analytical models show that without this feature, software scaling would idle tensor cores 40\% of the time, destroying performance gains. We are the first to leverage and demonstrate the necessity of this hardware capability for practical emulation.
\item Power efficiency quantification: First comprehensive power analysis showing 40\% GFLOPS per Watt improvement for scientific computing.
\item Hybrid Performance Improvements: We show significant performance improvements over native FP32 SGEMM for many sizes, and a hybridized-approach (first to use a hybridized-approach for FP32 emulation) which selects the fastest method. This allows users to simply call SGEMM and get the best performance.
\item Significant accuracy gains in real-world applications: Using BF16x9 instead of BF16x6 (or any of the other FP16 variants) yields more accurate results, and we show this in several numerically challenging applications across a wide range of disciplines.
\end{enumerate}


\section{Related work} \label{sec:related}

Baboulin et al \cite{baboulin2009accelerating}
focused on getting FP64 accuracy from FP32 computation on dense and sparse matrix linear solution solvers (such as $Ax=b$), and extended this to quad-precision. This work sought to exploit machines where FP32 ran more than twice as fast as FP64, so doing more than twice the work in FP32 might be justified. Their algorithm used iterative-refinement in a linear systems solve as a basis. 
Each time one does an iterative refinement scheme, it must be analyzed relative to the problem it is trying to solve. Additionally, there are issues associated with the conditioning of the problem, and if it is too ill-conditioned, the lower precision solution may have insufficient accuracy and thus the method will fail. To get the best results in iterative refinement, you also have to find the residuals in the desired higher precision.

But this is only the first method of exploiting lower precision. The other method is emulation, which replaces some higher precision calculations (ideally, the majority of them) with lower precision counterparts. 

Henry et al.~\cite{henry2019leveragingbfloat16artificialintelligence} split a single precision FP32 number into three BF16 values for the purpose of emulation, similar to our implementation. But there, the scalars were implicitly embedded into the exponents, making the matrix multiplies slightly further apart and not allowing for our recent Blackwell optimizations. The conversion was done through a routine that converts a FP32 number into a BF16 value (rounded), and then computes the error in FP32, uses that rounded error for the second significant BF16 value, and uses the accumulated error of the first two BF16s to compute the final BF16 value. Since FP32 has twenty-four mantissa bits (including the implicit bit) and a triplet of BF16s has $8\cdot3=24$ mantissa bits as well, it is possible to do this conversation without losing accuracy as the range is the same. Again, the difference lies in a minor way of handling the scalars, but it was necessary to use our current method to exploit the Blackwell optimizations.

In other words, $x = s_0*b_0 + s_1*b_1 + s_2*b_2$ where $x$ is a FP32 number, $b_i$ are BF16 numbers, and we focus on the case where $s_i$ are power-of-two scalars, including possibly one (since the range of $x$ and $b_i$ are the same, one can imagine each $s_i=1$, but since $b_0$ covers the first 8 bits and $b_1$ covers the second 8 bits, one can also imagine $s_0=2^0=1$ and $s_1=2^{-8}$ and $s_2=2^{-16}$). Implicitly, we could also choose to embed the scalars into our BF16 numbers, and assume $s_0=s_1=s_2=1$. Those are the two most practical cases ($s_0=1,s_1=2^{-8},s_2=2^{-16}$ or $ s_0=s_1=s_2=1$).

We will discuss this further in future sections.

When computing $d=c + a*b$ for scalars, where $a$ and $b$ are BF16, we assume that $c$ and $d$ are FP32 and that the addition and accumulation are done in FP32. There are variants that keep everything in BF16 arithmetic, but since hardware exists on NVIDIA GPUs to do the accumulation in FP32, and we are emulating FP32, we assume that all additions (except those within a BF16-GEMM) are done in FP32 arithmetic. Only the multiplications, including matrix-matrix multiplication, use BF16 arithmetic.

In order to study how the accuracy of an emulated SGEMM compares to native FP32 hardware, it is necessary to first review related work on the accuracy of SGEMM. SGEMM computes $C \leftarrow \beta C + \alpha \text{op}(A)*\text{op}(B)$ where $\text{op}()$ can represent the matrix in a standard fashion, or in a transposed fashion. A transposed matrix can also include a conjugate transpose, but for simplicity here, we assume the data is real. If $\text{op}(A) \in \mathbb{R}^{mxk}$, and $\text{op}(B) \in \mathbb{R}^{kxn}$ then $C \in \mathbb{R}^{mxn}$. 
Since how the data is accessed should not bear on accuracy, for the rest of this paper, we shall assume that the non-transposed case is used. One might also observe that if there is high cancellation error so that $C_{(i,j)}$ is small, but the original value on input for a non-zero beta case is larger (by a magnitude more than 24 mantissa bits can hold), that any inaccuracies in the GEMM will be absorbed by the final addition. So, from an accuracy perspective, $\alpha=1$ and $\beta=0$ are the most interesting cases. That is, we wish to focus on comparing our emulated SGEMM for the simplified case of $C=A*B$.

In the error analysis of $C=A*B$, notice that this is really $mn$ separate dot products, so it suffices to study the error analysis of a single native FP32 dot product and how it compares to its emulated counterpart. Suppose $x$ and $y$ are vectors of length $k$. The absolute error of the dot product between $x$ and $y$ is given by $ | x^T y - fl(x^T y) |$ where $fl()$ represents the floating-point computation in question. If $\mu$ is the machine-precision ($2^{-24}$ for FP32), then we have via the triangle inequality:

\begin{equation}
| x^T y - fl(x^T y) | \le k * \mu * |x|^T |y| + O(\mu^2)
\end{equation}

If we let $f(y)=x^T y$, and $J(y)$ be the derivative of $f$, then $J(y)=x^T$ and this yields the condition number of the dot product, 
denoted $\kappa (x,y)$, which is given as follows (assuming $x^T y \ne 0$):
\begin{equation} \label{eq:conddot}
\kappa (x,y) = \frac{|| x || * || y ||}{|x^T y|} = \text{sec}(\theta)
\end{equation}
where $\theta$ is the angle between the vectors, so near orthogonal vectors have the worst conditioning. In particular, if $x$ and $y$ are normalized, then these formulas can be further simplified. Also, unlike linear systems condition numbers, where there is one condition number for the entire linear system of $Ax=b$, we now have the distinct case of $mn$ independent condition numbers. 

Finally, we see that the relative error for a normalized dot product is going to be bounded above by:
\begin{equation} \label{eq:relerrdot}
Relative Error \le k * \mu * \kappa(x,y)
\end{equation}
plus some very small terms that we can largely ignore. In practice, the relative errors are better, usually in the range of 4 or 5 bits better. That is, $\mu = 2^{-28}$ in Equation~\ref{eq:relerrdot} might fit the observed FP32 relative errors better. But a worst case scenario can be created that achieves the maximum theoretical error.

Cancellation errors are clearly represented well in these Equations~\ref{eq:conddot} and~\ref{eq:relerrdot}. Absorption errors are not as well represented. But in practice, errors in floating-point while doing a dot product tend to come from cancellation errors. Absorption errors may become prominent in the final addition (when the GEMM has a $\beta \ne 1$), but analysis focuses on cancellation errors. 

In \cite{henry2019leveragingbfloat16artificialintelligence}, an analysis was done on $fl(x^T y)$ errors versus the floating-point error of breaking each FP32 term into 3 BF16s, and instead of having one FP32 product, having nine BF16 products. That is, if $x=x_0 + x_1 + x_2$ (ignoring the potential scaling values $s_i$ discussed earlier) and $y=y_0 + y_1 + y_2$, then $x^T y$ is the sum of nine terms. 
One could select the six most significant products and drop the three potentially smaller products to get only six multiplies. The decomposition of the data is the same, but the work and accuracy of the results differ. We denote a GEMM based on 9 BF16 matrices as BF16x9 SGEMM (or ``emulated SGEMM") and we wish to compare fl(BF16x9 SGEMM) and the floating-point calculation of the native FP32 GEMM (or fl(SGEMM)). In \cite{henry2019leveragingbfloat16artificialintelligence}, the worst case component-wise relative error in $C=A*B$ done via nine BF16 multiplies, or BF16x9 SGEMM, was approximately the same as the native SGEMM, but it may be slightly worse under the right conditions. 

The accuracy of BF16x9 has been analyzed in related work, but before we share those final results, note that a nine product BF16x9 SGEMM can be implemented in a number of different ways. One can use multiple FP32 accumulators, add up values in different orderings, and round or truncate differently. Final results about accuracy can not be stated precisely without getting into significant implementation details.
Also, the final results can never show a generic emulation always beating SGEMM, because SGEMM might get lucky and round to exact answers. There's no magic bullet for ``emulation always beating FP32." For this reason, in later sections, we will try to show as much comparable data as possible.

The result that emulation of FP32 with triplets of BF16 is more accurate more often than not is intuitive. The algorithm decomposes a floating-point number into pieces and doing lots of smaller calculations, but always using FP32 arithmetic when gluing these pieces back together. Intuitively, this should yield greater accuracies.

In the recently mentioned \cite{henry2019leveragingbfloat16artificialintelligence}, BF16x6 was used to emulate FP32, where the three least significant multiplies are dropped from the BF16x9 emulation. While this on average behaves quite similarly to BF16x9, even for dot products that are highly ill-conditioned, the worst case error tends to be slightly more accurate for BF16x9. Since BF16x6 and BF16x9 share the same slicing and memory movement for the BF16 matrices, doing only two-thirds of the work does not yield the full implied savings.
Using the Blackwell tensor cores, however, we still achieve great performance results, even using the 9 multiplies, and this method leads to a way to deal with denormals. The smallest possible exponent in FP32 is -126, which means the smallest power of two representable in FP32 is -149 (because FP32 has 23 explicit mantissa bits). If one knows the range of the exponents lies between [-110,111], one might be able to represent both small and large numbers correctly in BF16x6, otherwise we have a robust solution for BF16x9- all made possible because of Blackwell tensor cores.

We do more than a mere CPU version of this emulation, we bring a GPU implementation to the table. And, as mentioned, ours is library-ready, robust, and handles denormals.

There are other methods of doing emulation for higher precision. Ozaki et al~\cite{10.1007/978-3-030-50743-5_12} and Schwarz et al~\cite{nv_dgemm_guarantee26} have studied breaking up a higher precision mantissa into multiple smaller pieces, sometimes leveraging fixed-point as well as floating-point. There are details in the implementation, but overall we see the same kind of expansion and selection of a given number of multiplies from this. That is, $A = s_0 A_0 + s_1 A_1 + \dots\ + s_m A_m$ for some higher precision $A$ and lower precision $A_i$ with potential scalars $s_i$, and $B$ is also expanded out in terms of $n$ lower precision $B_j$, and then the product is $mn$ products (or only the most significant ones) of $A_i B_j$ as $i=1\dots m$ and $j=1 \dots n$. In \cite{parikh2024cascadinggemmhighprecision}, this kind of polynomial was referred to as a Cascading GEMM, a term applicable whenever there are a sequence of $A$s that are all multiplied by a sequence of $B$s. There is also a notion of Batched-GEMM where a user has a set of $A$ and $B$ matrices and needs to find a product, but where there is no implicit reuse assumed. That is, one computes $A_0*B_0$ but never $A_0*B_3$. In Batched-GEMM, each multiply is independent of the others.  Thus, if $A_0$ is used multiple times it must be listed multiple times. While in Cascading GEMM, one assumes that one needs to at least approximate each $A_i * B_j$ for all $i$ and $j$.

Many papers also follow this same Cascading GEMM methodology. One could even emulate FP32 using 8 bit fixed-point, but this would introduce range issues that BF16x9 does not have. So, for reasons related to both performance and range, using BF16 in emulating SGEMM, is highly motivated. 

In \cite{ootomo2022recovering}, they also work with FP32 accuracy through splitting into lower precision pieces, specifically exploiting FP32 accumulation. In their case, they used FP16x4 and TF32x3. They claim that TF32x3 is as accurate as native FP32.

In our work, we see performance improvements significant enough on the GPU to use this emulation technology in practice. 

\section{Algorithmic advancements}\label{sec:contrib}

While previous works may contain prototyped experiments, we do the full emulation of BF16x9 and show performance improvements for a library-ready implementation that includes denormals. This is a significant contribution. We also extend numerical tests done in the past to illustrate the value of our technique. We do this not just for uniformly random data, but for a wide range of exponents and we have a special generator for $A$ and $B$ that produces $mn$ dot product condition numbers that average around the desired input (and most of the condition numbers in practice get close to the desired average condition number). 

Because of these contributions, now AI hardware can be used in scientific computing in a robust environment. 

We present two categories of accuracy studies performed on FP32 matrix multiply emulation. We will first examine the accuracy of the matrix multiplication routine in isolation, comparing FP32 matrix multiplication (SGEMM) using the BF16x9 emulation method to both SGEMM using native FP32 hardware, and DGEMM, utilizing native FP64 hardware.  We will go on to evaluate the accuracy characteristics of three scientific applications, utilizing each of these matrix multiplication implementations.

\section{BF16x9 emulation algorithm and special values (denormals, NaN, Inf)}\label{sec:algo}

\begin{figure} \label{fig:1}
    \centering
    \includegraphics[width=1.0\linewidth]{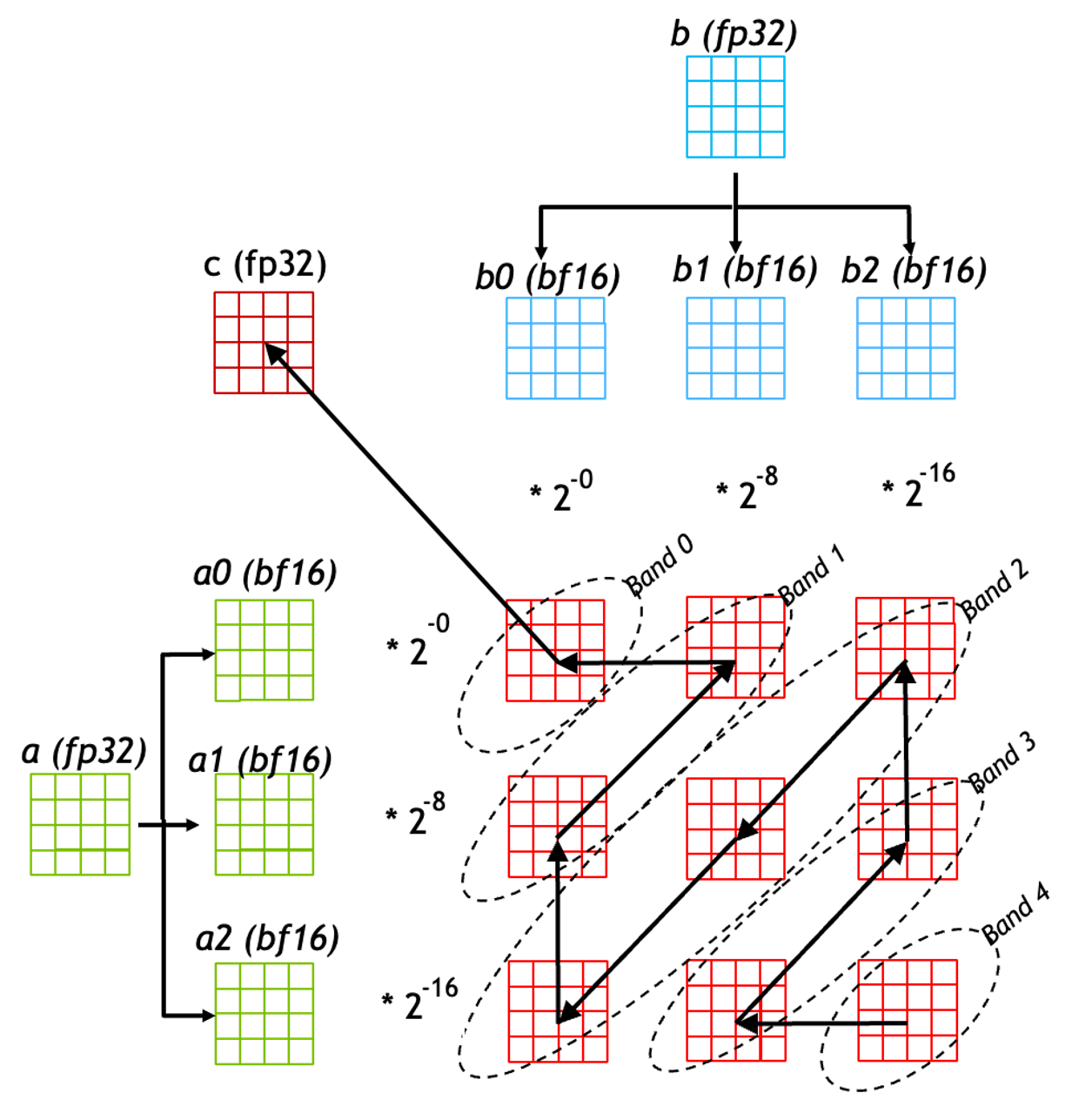}
    \caption{Illustration of how matrix multiply is implemented with Tensor Cores on a 4-by-4 example. The \(a\) and \(b\) matrices are decomposed into three BF16 matrices each. Scaling factors for each row and column are also shown. The resulting nine products are accumulated in FP32 along five bands, as shown by the arrows, to minimize rounding errors. The final result is added to the \(c\) matrix.}
    \label{fig:matmul1}
\end{figure}

We use the fused multiply-add (FMA) operation \(d = a \cdot b + c\) to illustrate the FP32 emulation algorithm, using BF16 tensor cores. Each FP32 number is decomposed into three BF16 numbers using the so-called \textit{elementwise-place splitting} method \cite{ootomo2022recovering} in the following manner:
\begin{equation} \label{eq:1}
    \begin{gathered}
a = a_{0} + 2^{-8} \cdot a_{1} + 2^{-16} \cdot a_{2}\\
b = b_{0} + 2^{-8} \cdot b_{1} + 2^{-16} \cdot b_{2}
    \end{gathered}
\end{equation}
Scaling at each split step in the decomposition is used to improve accuracy, where the scaling factor is equal to the number of mantissa bits (including the implicit bit). Using Equation \ref{eq:1}, we can then implement the FMA for scalars as follows:
\begin{equation} \label{eq:2}
    \begin{split}
d = & a_{0} \cdot b_{0} + 2^{-8} \cdot a_{0} \cdot b_{1} + 2^{-16} \cdot a_{0} \cdot b_{2}\\
  + 2^{-8} \cdot & a_{1} \cdot b_{0} + 2^{-16} \cdot a_{1} \cdot b_{1} + 2^{-24} \cdot a_{1} \cdot b_{2}\\
  + 2^{-16} \cdot & a_{2} \cdot b_{0} + 2^{-24} \cdot a_{2} \cdot b_{1} + 2^{-32} \cdot a_{2} \cdot b_{2} + c
    \end{split}
\end{equation}
In practice, we implement Equation \ref{eq:2} using Tensor Core operations for matrix multiplies, as shown in Figure \ref{fig:matmul1}. Since we are performing nine times the FMA operations, the maximum computational throughput of this method is \(\frac{1}{9}^{th}\) of the Tensor Core throughput for BF16.

In order to reach close to the peak theoretical throughput of BF16x9 emulation while delivering the accuracy of FP32, as we accumulate anti-diagonals, we leverage hardware support for integrated scaling, available on the most recent datacenter Blackwell GB200 GPUs.  On these systems, the BF16 tensor cores are equipped with hardware support for scaling, allowing us to avoid the significant overheads we would experience were we to perform this scaling in software. This scaling is performed through the use of the \texttt{tcgen05.mma} instruction and providing an input to \texttt{scale-input-d}, see \cite{nvid24b}. Our projections, based on an analytical throughput model, indicate that, on GB200, applying scaling and accumulation frequently enough to maintain the desired level of accuracy, without hardware support, could leave the tensor cores idle as much as 40\% of the time, a significant impact, illustrating the importance of this feature.

\begin{figure}
    \centering
    \includegraphics[width=0.9\linewidth]{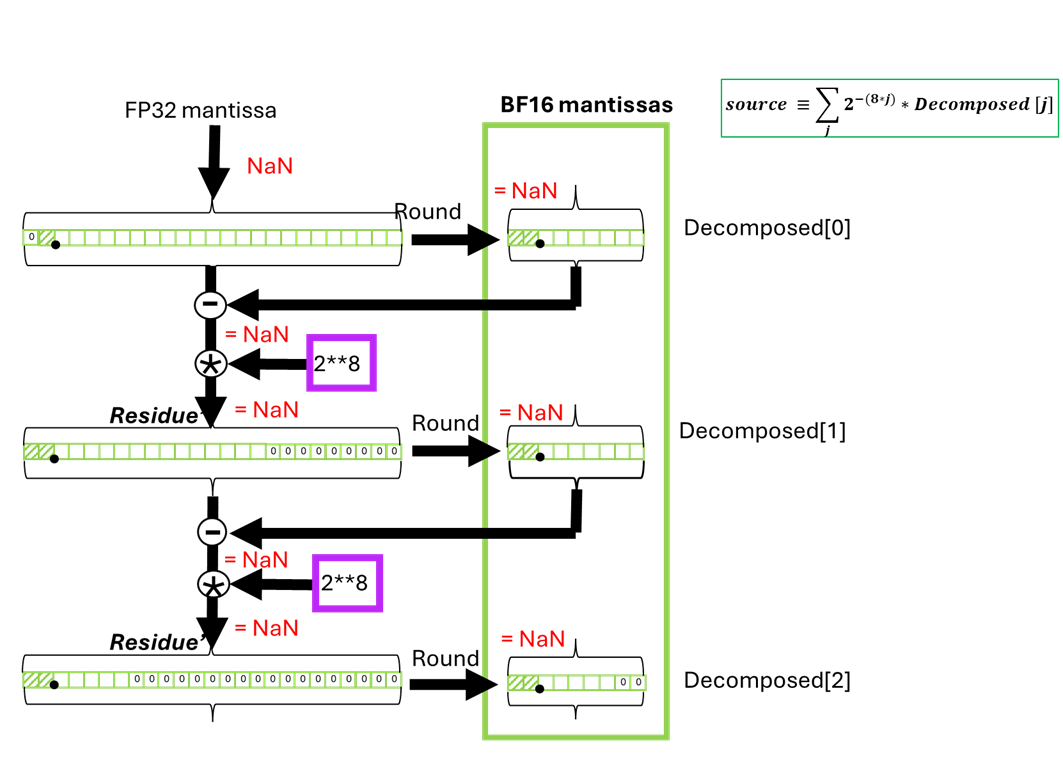}
    \caption{NaN propagation during decomposition}
    \label{fig:decompose-nan}
\end{figure}

\begin{figure}
    \centering
    \includegraphics[width=0.9\linewidth]{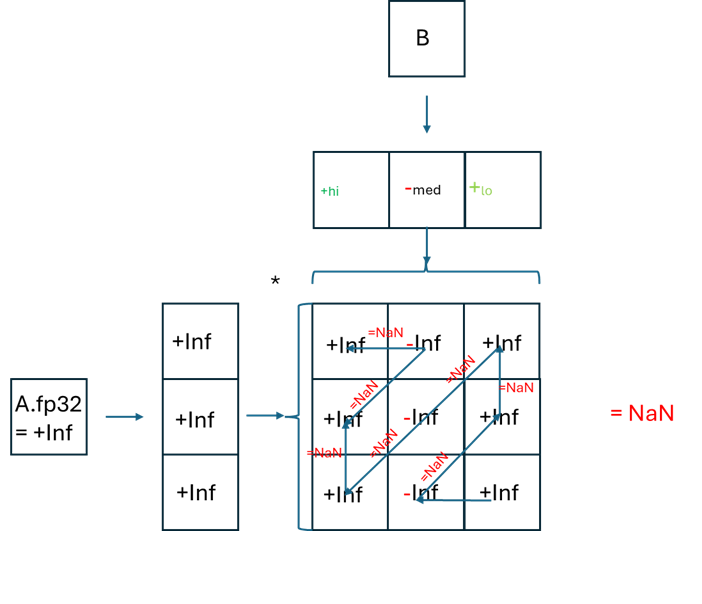}
    \caption{Spurious NaN generated during infinity * x(finite) multiplication}
    \label{fig:enter-label}
\end{figure}

Special signed \texttt{Inf} and \texttt{NaN} input FP32 values deserve special consideration. NVIDIA FP32 subtract and multiply and half-precision rounding intrinsics involved in the decomposition, as well as the Tensor Core units accept and propagate the \texttt{NaN}s to all data-dependent outputs in accordance with the usual IEEE rules, which is the desired behavior. 
Figures~\ref{fig:decompose-nan} and ~\ref{fig:enter-label} illustrate problems with \texttt{NaN} propagation and spurious \texttt{NaN} issues.
When signed FP32 infinity encodings in the inputs are expected, there are two options:

a) Decompose FP32 Infinity into triplets of signed BF16MAXFINITE. These recompose back to FP32MAXFINITE, so this behavior is functionally equivalent to the consistent behavior of saturating FP32INF in the input multiplicand matrices to FP32MAXFINITE first, then performing the decomposed FP32 GEMM on the finite values normally in accordance with Eq.~\ref{eq:2}.

b) Decompose FP32 signed infinity into triplets of signed BF16INF. While it is an added cost to the ALU to perform the \texttt{extrq} override, which does result in more energy consumption per unit of computation, it is not on the critical path for performance. A bigger issue with this option is the fact that a finite FP32 value in the opposing multiplicand matrix may decompose into a finite BF16 of opposing signs. While a standalone Tensor Core  matrix multiply does handle signed infinity in accordance with the relevant IEEE rules, in such a common scenario the multiply-accumulation of same-sign BF16 infinity triplet multiply-accumulated with the different-sign finite BF16 triplet generates nine intermediate infinity products of opposing signs. Therefore, the partial dot product being formed resets to NaN after the first occurence of opposite-sign BF16 Infinity product. 

Therefore, we opt for the consistent behavior provided by the option a) 

In a library implementation, propagating error conditions is vital, and we utilize a patching framework to update elements which should be an infinity or Nan.  The results may deviate from IEEE-754 error conditions under some scenarios. In particular, infinities may propagate as NaN but we guarantee that an error condition is propagated with minimal performance overhead.

\section{Numerical Accuracy Studies}\label{sec:numacc}

\begin{figure}
\centering
\includegraphics[width=1.1\linewidth]{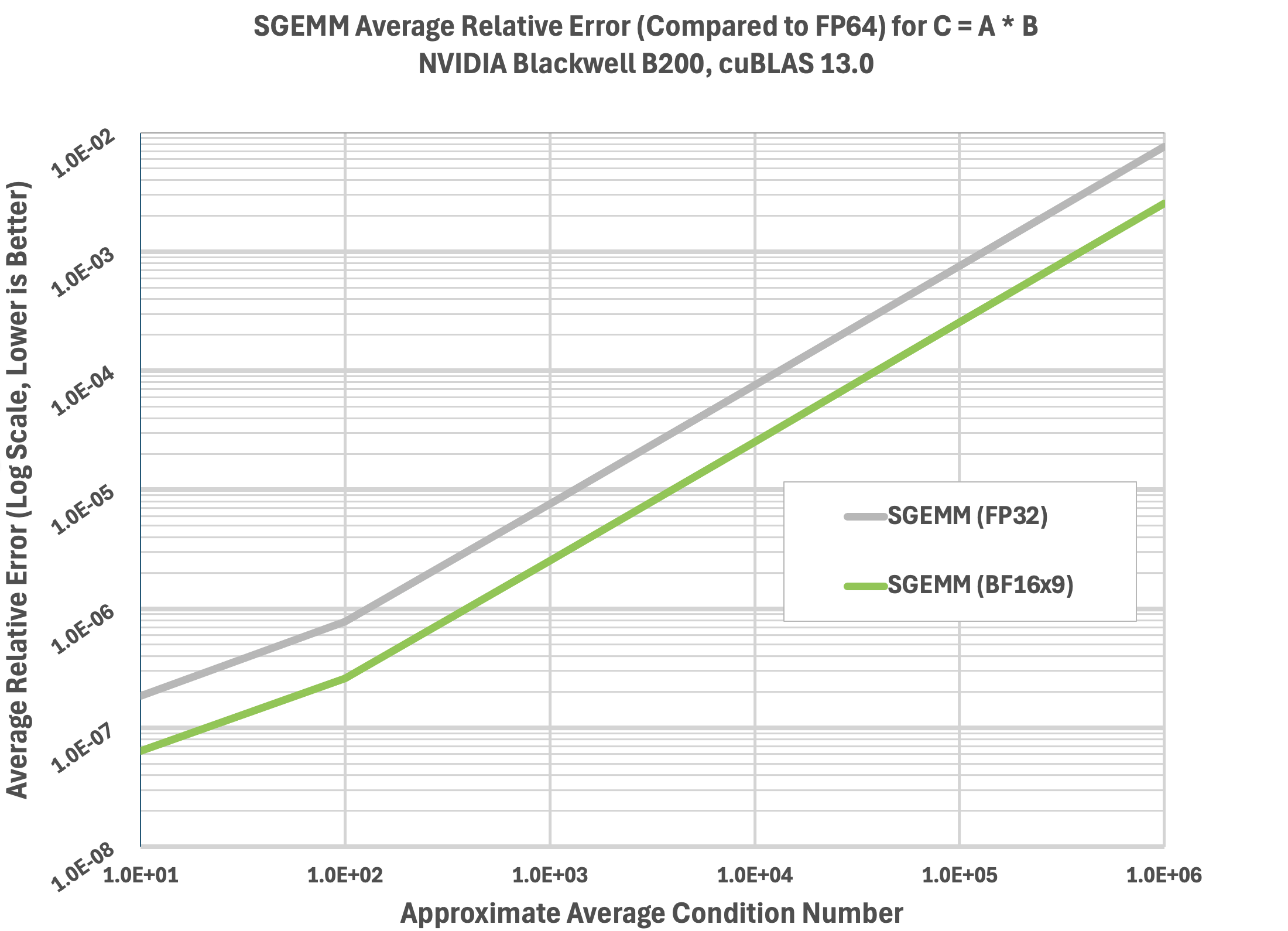}
\caption{32-bit SGEMM Average Relative Errors for $C=A*B$. The two lines are native FP32 SGEMM and emulated BF16x9 SGEMM average component-wise ($160^2$ elements) relative error as the average condition number for the data varies. DGEMM was used as a reference and averages over 10000 different 160x160 matrices were used.}
\label{fig:conditioning1}
\end{figure}

\begin{figure}
    \centering
    \includegraphics[width=1.0\linewidth]{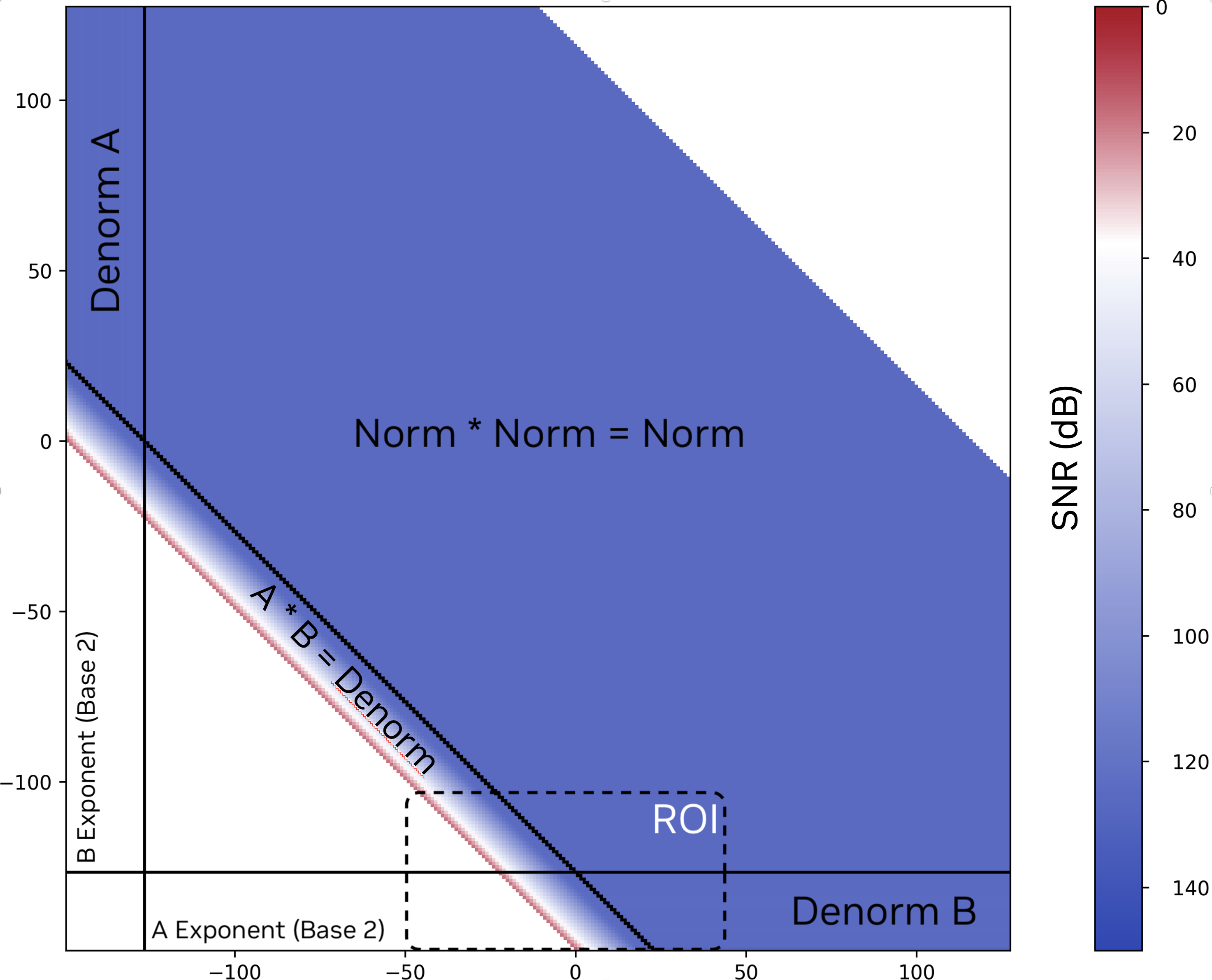}
    \caption{Numerical accuracy results comparing FP32 to FP64 FMA based implementations for multiplication of two matrices A[512x1024] and B[1024x2048] which serves as a reference for our BF16x9 implementation. The heatmap shows the SNR in dB as defined in \eqref{eq:RMS} and \eqref{eq:SNR} for varying exponent combinations. The region of interest (ROI) encompasses the four combinations of normal and denormalized multiplications. 
    }
    \label{fig:accuracy1}
\end{figure}

In this section we present numerical accuracy tests that compare the results of our emulation scheme with both standard FP32 and FP64 matrix multiplies. The accuracy of emulation methods can be impacted by both conditioning and exponent range. Figure \ref{fig:conditioning1} shows how the condition number for each dot product on average impacts the final relative errors. Figure \ref{fig:accuracy1} shows how the exponents of the two input matrices, $A$ and $B$, are varied along the two axes, resulting in various normal and denormalized (\textit{aka subnormal}) number combinations.

In Fig.\ref{fig:conditioning1}, we generate thousands of pairs of matrices, $A$ and $B$, whose average dot product condition number ranges from 1.0e+1 to 1.0e+6, and for each of these pairs, compute the average component-wise relative error (as compared to DGEMM) of both native SGEMM and emulated SGEMM. Suppose our desired condition number is $\delta$ ($\ge 1.0$). The matrices are generated in reverse, starting by generating a $C$ matrix with randomly chosen values in the range $\pm [.9*\frac{1}{\delta},1.1*\frac{1}{\delta}]$ where the sign is computed randomly, and a random location of $C(:,j)$ in every column $j$ has a value near one (all other values are near zero when $\frac{1}{\delta}$ above is chosen small). Then each column of $C$ will have a norm near one, especially if $\frac{1}{\delta}$ is a small positive number. We also generate a random orthonormal matrix, which we can diagonally scale for further randomness, named $A$, and let $B=A^T * C$. Now $C=A*B$ in exact arithmetic and the norm of every column in $B$ should be near 1, and the norm of every row of $A$ should be near one, modulo any diagonal scaling. This means that most of the dot product condition numbers for $C$ will average $\delta$. Because $O(n)$ of the quadratic dot products will have condition 1, the average will probably be shy of $\delta$ for larger matrices, and we have seen this in practice. We then ran ten thousand pairs of $A$ and $B$, for various condition numbers and choices of $\delta$, and examined the relative errors, as compared to DGEMM. Each data point represents an average over $10000*160^2$ numbers. The results show a distinct average advantage for emulation. The same can be done with the average maximal relative error as well, but that chart is not included for space. It is also worthwhile noting that most of the BF16x9 SGEMM cases had a higher relative accuracy compared to their native FP32 counterparts (usually over 60\% of them), so it was not that this average is unbalanced by a few great cases, but that overall the BF16x9 SGEMM appeared superior for these tests.

\begin{figure}
    \centering
    \includegraphics[width=1.0\linewidth]{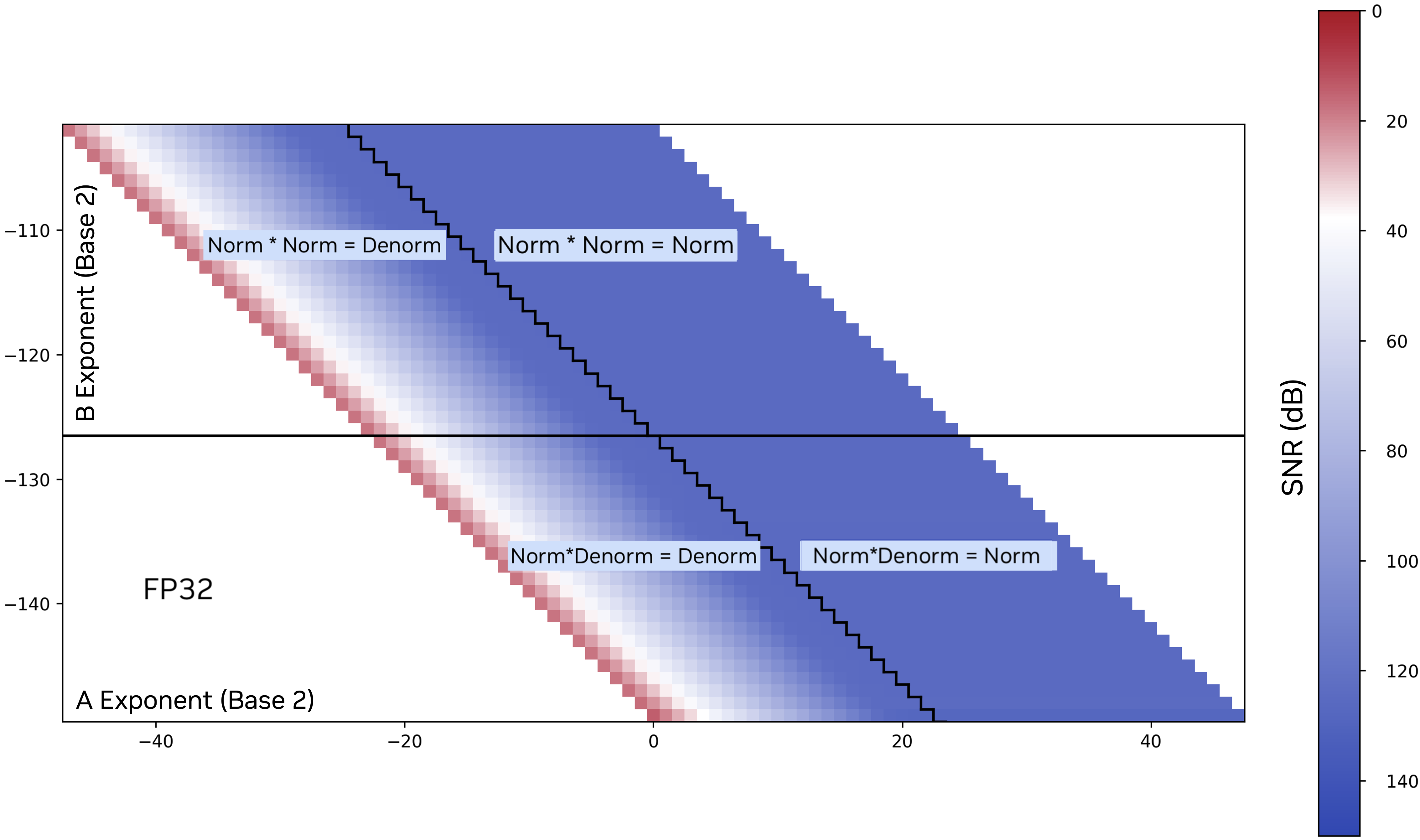}
    \includegraphics[width=1.0\linewidth]{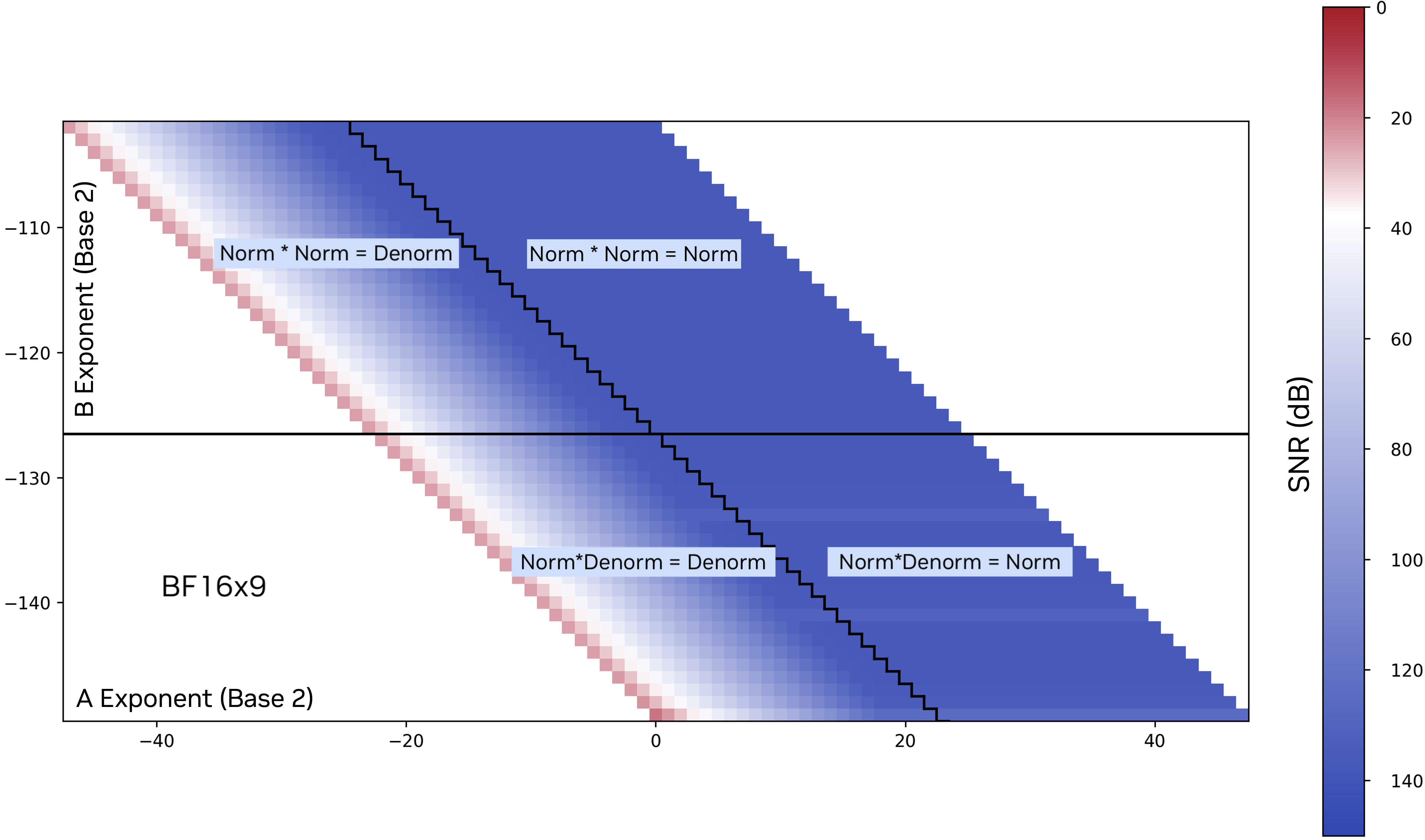}
    \caption{Comparison of the numerical accuracy heatmaps of FP32 and BF16x9 implementations to FP64 in the ROI shown in Figure \ref{fig:accuracy1}. The results clearly show that errors relative to FP64 are lower for BF16x9 than for FP32.
    }
    \label{fig:accuracy2}
\end{figure}

The error shown in Fig.~\ref{fig:accuracy1} and Fig.~\ref{fig:accuracy2} is calculated as the root mean square (RMS) error (Equation \ref{eq:RMS}) and is plotted as the signal-to-noise ratio (SNR) in decibels (dB) (Equation \ref{eq:SNR}).

\begin{equation} \label{eq:RMS}
    \begin{gathered}
RMS = \sqrt{\frac{\Sigma_{i,j}(Result_{i,j} - Result^{FP64}_{i,j})^{2}}{\Sigma_{i,j}(Result^{FP64}_{i,j})^{2}}}\\
    \end{gathered}
\end{equation}

\begin{equation} \label{eq:SNR}
    \begin{gathered}
SNR = -20.0 \cdot \log_{10}(RMS)\\
    \end{gathered}
\end{equation}

\section{Impact in Scientific Computing: Numerical Accuracy}\label{sec:impact}

To validate the accuracy, performance, and efficiency of the cuBLAS implementation of SGEMM implemented via the BF16x9 algorithm, we conducted both standard benchmarking and experiments with scientific applications using real-world data.  This section details the accuracy studies in the domains of weather forecasting, quantum circuit simulation, quantum chemistry, and condensed matter physics.  These specific applications are widely accepted as amenable to FP32 compute for their arithmetically intense components and our results verify that SGEMM using BF16x9 emulation results in errors that are comparable to, and usually superior to, those seen when using standard FP32 SGEMM.

\subsection{Weather Forecasting} \label{subsec:numacc_exp_weather}

\subsubsection{Description}
Spherical harmonics are commonly used to represent data in weather and climate applications such as IFS and Arpege. ecTrans is a library developed by ECMWF to perform transformations between the grid-point and spectral representation of scalar data such as temperature, and vector data such as velocities. Velocities are represented as Vorticity and Divergence in spectral space. The correctness and accuracy of the transformations can easily be evaluated by tracking the error distribution of a real field by applying a series of forward and backward transformations. While the error grows exponentially at reduced precision (TF32), it usually grows linearly at single precision.
\subsubsection{Numerical Results} We ran 1000 consecutive forward and backward transformations of the spectral transform onto real data fields extracted from an actual simulation and tracked the error distribution of velocities (U and V) and a temperature, using FP32 (precision operationally used at ECMWF for the daily forecasts) and BF16x9-emulated FP32 arithmetics.

\begin{figure}[h]
    \centering
    \includegraphics[width=\linewidth]{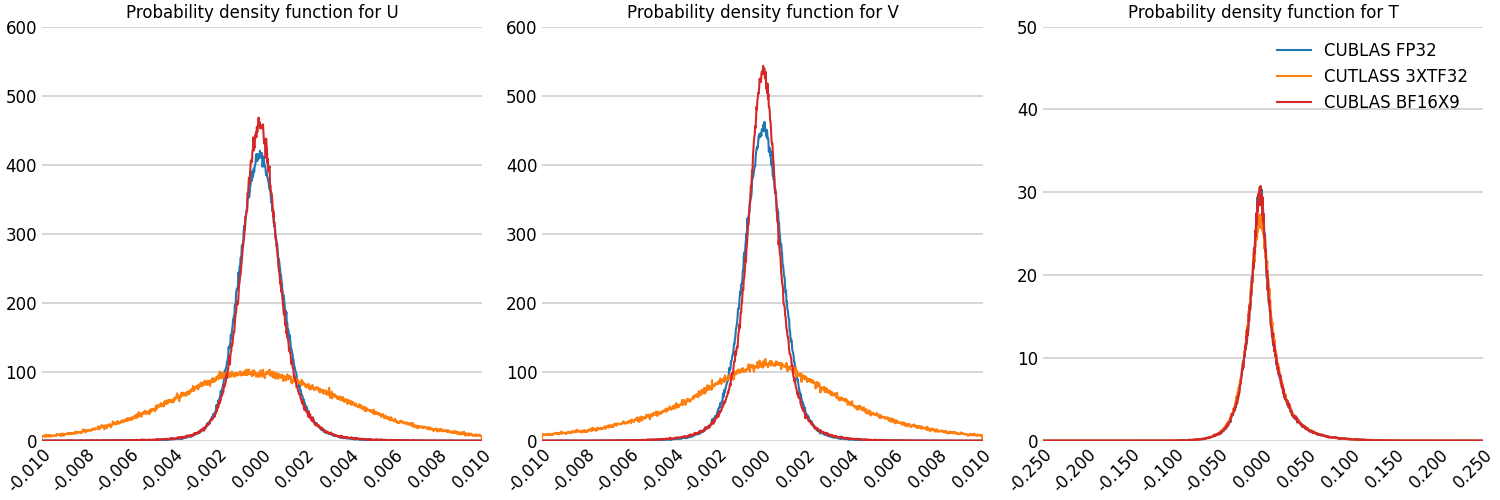}
    \caption{Probability density of the error after 1000 forward and backward iterations at low resolution using TCo399 (~25 km global resolution), showing equal precision for BF16x9 compared to FP32, while 3xTF32 shows worse accuracy for velocity fields.}
    \label{fig:weather1}
\end{figure}

\begin{figure}[h]
    \centering
    \includegraphics[width=\linewidth]{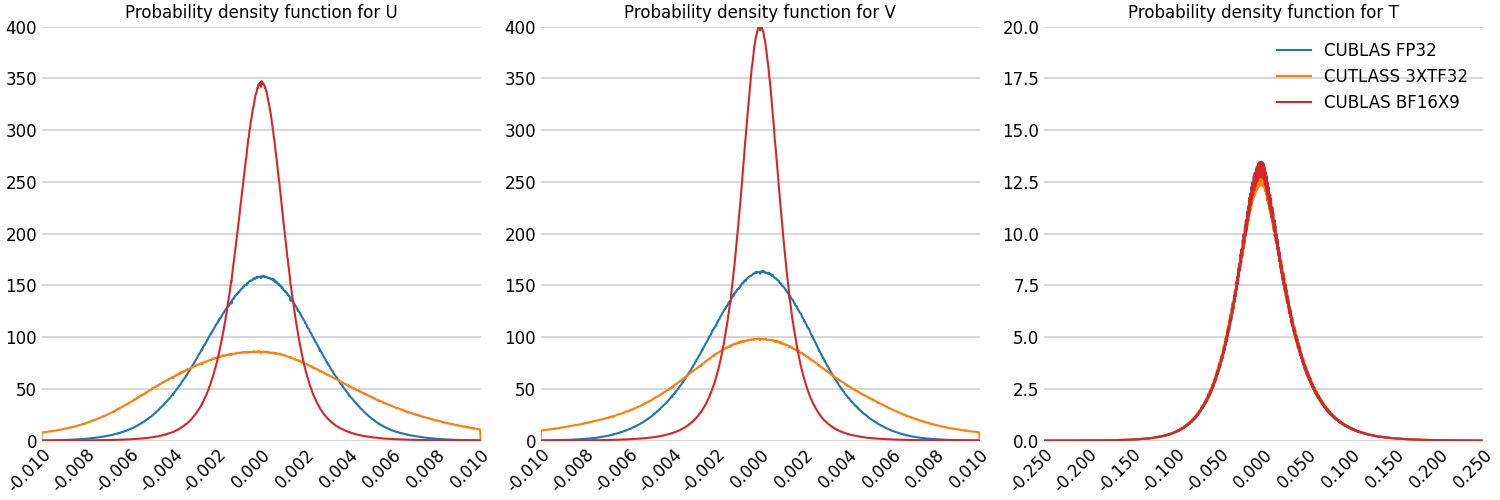}
    \caption{Probability density of the error after 1000 forward and backward iterations at high resolution using TCo3999 (~2.5 km global resolution), showing superior performance of BF16x9 compared to other math precisions.}
    \label{fig:weather2}
\end{figure}

The results in Fig.~\ref{fig:weather1} show that the error distribution of the temperature field behaves the same for BF16x9-emulated SGEMM as for native FP32 SGEMM. The error growth after 1000 iterations in general is below 0.1 degree, which is neglectable. For temperature, the errors also grow very small, but for BF16x9-emulated SGEMM the error grows even slower than for native FP32. The difference in behavior between temperatures and velocities is likely due to the extra computations required to transform velocities into vorticity.

\subsection{Quantum Computing}

\subsubsection{Description}

Quantum circuit simulations are vital for quantum algorithm design and verification. The current NISQ quantum processors have already surpassed the limits of the state-vector-based simulation approaches in terms of the number of qubits. As a remedy, tensor-network methods have become an indispensable tool for implementing more scalable quantum circuit simulators, in particular those used for probing the quantum supremacy boundary based on the random quantum circuit simulations \cite{Arute2019, Villalonga2020, Nguyen2022, Gray2018, Pan2022}. These random quantum circuits are designed to be hard to simulate on classical computers. The tensor-network-based simulation proceeds by converting a quantum circuit into a tensor network, followed by contracting all tensors of that tensor network to produce a slice of the final wave-function tensor. 
The corresponding wave-function tensor is characterized by a rather flat distribution of tensor element values. In particular, the majority of the elements of the wave-function tensor of the 53-qubit random quantum circuit used to validate Google's Sycamore quantum chip \cite{Arute2019} have values in the range $10^{-10}$ - $10^{-8}$. Previously, it was shown that single-precision (FP32) simulations are numerically accurate whereas reduced precision simulations (TF32) introduce a noticeable error when simulating such random quantum circuits \cite{Nguyen2022}. Here we extend that study by analyzing the numerical accuracy of FP32 arithmetics emulated via the BF16x9 emulation scheme.

\subsubsection{Numerical Results}

We simulated the 53-qubit random quantum circuit used for validating the Sycamore quantum chip with 12 layers of random gates and computed the probability amplitudes for a batch of 64 bit-strings. The corresponding simulation involves 649216 pairwise tensor contractions; 1.7\% of the tensor contractions accounts for 95\% of the total 0.83 PFLOP computational cost. The tensor network contraction path was determined by the cuTensorNet library \cite{Bayraktar2023} without heavy optimization. We generated multiple tensor network contraction paths to study the effect of the contraction path on numerical accuracy as well. All tensor contractions were computed with both FP64 (ground truth reference) and FP32 (typical precision used in quantum circuit simulations). Then we recomputed the same tensor contractions with the BF16x9-emulated FP32 arithmetic, where the emulation was only applied to those tensor contractions which map to a GEMM call with $k$-dimension $>= 16$ (these tensor contractions dominate the total computational cost). We used the RMS error (see Eq.\ref{eq:RMS}) of the computed probability amplitudes (wave-function tensor elements) as a measure of numerical accuracy.

The results summarized in Figure \ref{fig:QCS1} clearly indicate that numerical error introduced by the BF16x9-emulated FP32 arithmetic is about the same (slightly smaller) as the error of the native FP32 arithmetic, as compared to the FP64 baseline (first two candles). Moreover, the error is more uniform across the sample of the computed probability amplitudes in the former case. The 3rd candle demonstrates the numerical error of the BF16x9-emulated FP32 arithmetic with respect to the native FP32 arithmetic, and it is very small as expected, indicating a close proximity of the emulated and native FP32 results. Finally, the 4th candle demonstrates that the choice of different contraction paths does not significantly influence the numerical accuracy.

\begin{figure}[h] \label{fig:QCS1}
    \centering
    \includegraphics[width=\linewidth]{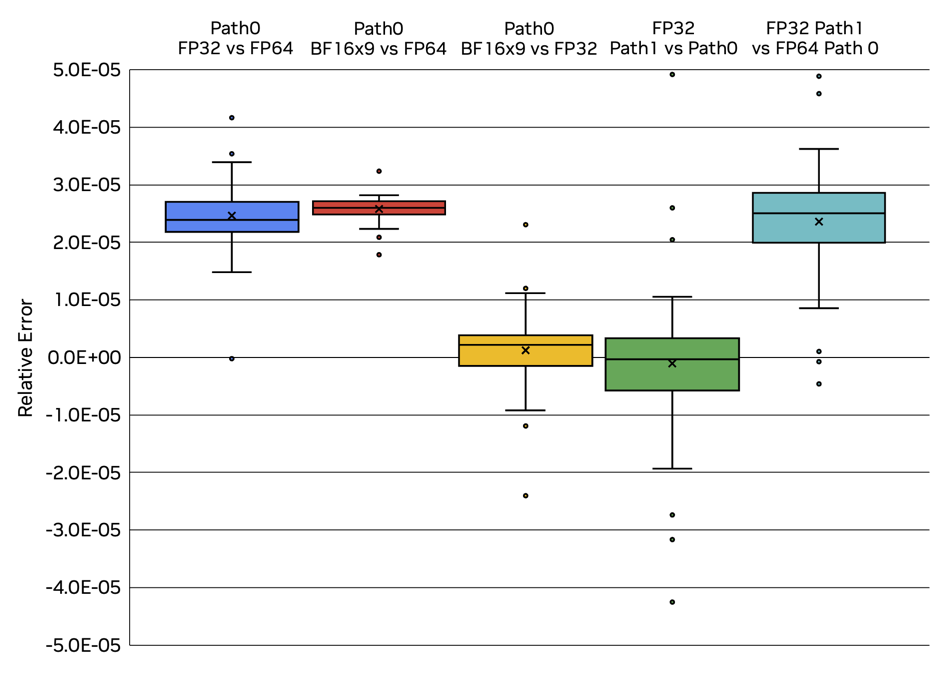}
    \caption{Box plot chart showing the RMS error for five different scenarios. From left to right: (1) Contraction path 0 computed with native FP32 arithmetic compared to the FP64 baseline; (2) Contraction path 0 computed with emulated FP32 arithmetic compared to the FP64 baseline; (3) Contraction path 0 computed with emulated FP32 arithmetic compared to the native FP32 baseline; (4) Contraction path 1 computed with native FP32 arithmetic compared to contraction path 0 computed with native FP32 arithmetic; (5) Contraction path 1 computed with native FP32 arithmetic compared to the FP64 baseline.}
    \label{fig:QCS1}
\end{figure}

\subsection{Quantum Chemistry}
\subsubsection{Description}
Coupled-cluster theory is used to model molecules to high precision, and is considered the gold standard in molecular modeling. In particular, the CCSD(T) variant, which is a correction to a fully-converged coupled-cluster singles and doubles (CCSD) calculation, is capable of achieving chemical accuracy. Additionally, the result of a coupled-cluster calculation may be used to train machine-learned interatomic potentials~\cite{Smith_Roitberg_2019} or density functionals~\cite{deepmind_21}, enabling high-precision simulations of large molecules. It is therefore desirable to perform CCSD calculations as efficiently as possible. In some challenging cases, the use of FP32 over FP64 may be a concern. Given that CCSD is an iterative method, it may be useful to perform the majority of iterations using FP32, and perform the final few iterations using FP64 if necessary. The use of precision lower than FP32 may not be acceptable in many cases, making emulated precision of FP32 an appealing approach.

Our implementation of CCSD follows the equations presented in~\cite{DePrince_Sherrill_2013}, without frozen-natural orbital truncation, based on the spin-free $t_1$-transformed equations introduced in~\cite{Koch_Helgaker_1994}. The density-fitted electron-repulsion integrals required in the CCSD equations are obtained using PySCF~\cite{pyscf}.
\subsubsection{Numerical Results}
\label{sec:ccsd-accuracy}
\begin{figure}[htbp]
    \centering
    \includegraphics[width=\columnwidth]{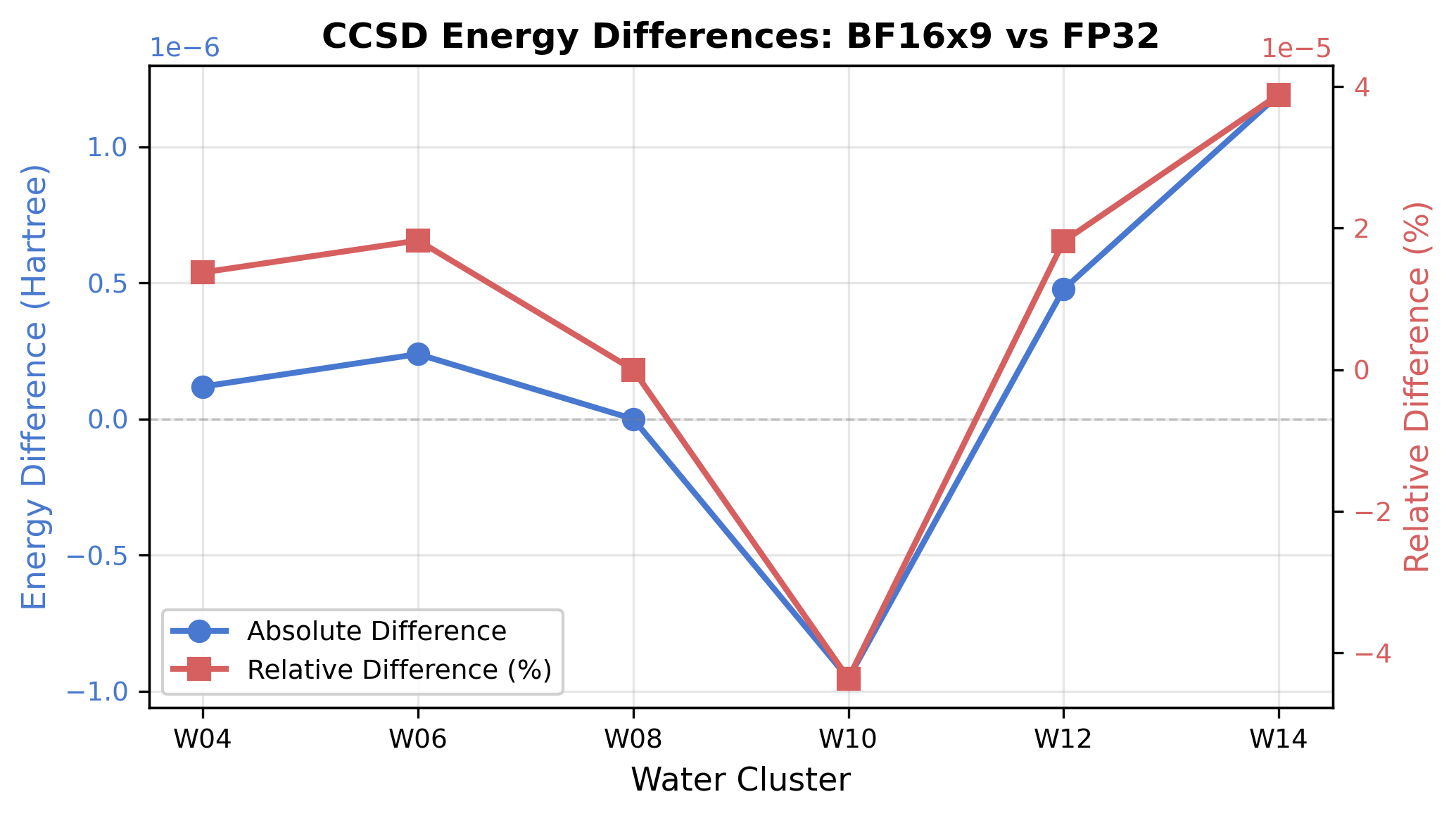}
    \caption{Comparison of converged CCSD energies for water clusters using native and emulated FP32.}
    \label{fig:ccsd_accuracy}
\end{figure}

We have performed CCSD calculations using native and emulated FP32 for all tensor contractions. We have checked the converged energies for a series of water-molecule clusters using the optimized geometries provided in~\cite{water_geometries}. We find that the results are nearly identical up to machine precision in all cases. Our results are shown in Fig.~\ref{fig:ccsd_accuracy}.

\subsection{Condensed Matter Physics}
\subsubsection{Description}

Tensor network theory is also broadly used in condensed matter physics for simulating quantum spin systems. In such simulations, the wave-function tensor of a quantum spin system is factorized as a tensor network with a chosen topology, for example, the matrix product state (MPS) \cite{Schollwock2011}, or the tensor tree state (TTS) \cite{Verstraete2013}, or some other topology. The factorization scheme significantly reduces both the memory size required for storing the wave-function and the computational cost required for computing the wave-function. Then the wave-function can be computed by minimizing the expectation value of the total energy operator by varying the elements of the tensors inside the tensor network representation of the wave-function, combined with some physical constraints, like normalization or symmetry. As in the case of quantum circuit simulations, the main numerical operation here is tensor contraction. However, since the expectation value calculation involves summation of many individual components, one can expect a more pronounced cancellation of error in this case. In addition to tensor contractions, the optimization algorithm involves the modified Gram-Schmidt (MGS) orthogonalization procedure~\cite{trefethen97}. Note that in our numerical results below we used the emulated FP32 arithmetic only in tensor contractions because the MGS procedure requires double precision for numerical stability whereas tensor contractions can run in single precision.

\subsubsection{Numerical Results}

As a representative benchmark, we have computed the ground state energy of the transverse-field Ising Hamiltonian for a chain of 16 spins using the tree tensor network factorization with the maximal bond dimension restricted to 16. The computation was performed via the ExaTN library \cite{Lyakh2022} with the cuTensorNet backend \cite{Bayraktar2023}. First we ran the full FP64 baseline simulation. Then we reran the simulation with native FP32 tensor contractions. Finally, we reran the simulation with the emulated FP32 arithmetic inside tensor contractions. The computed FP64 reference ground state energy is $-17.024189$ with the 2e-6 convergence tolerance. The computed native FP32 result is $-17.024197$ while the corresponding emulated FP32 value is $-17.024183$ (with the same convergence tolerance). Since the numerical difference between the native and emulated FP32 values is of the same order as the convergence tolerance, we conclude that the emulated FP32 arithmetic is as accurate in this case as the native FP32 arithmetic, and both are close to the reference FP64 value.

\section{Impact of BF16x9 Emulation on Performance in Scientific Applications} \label{sec:perf}

\subsection{Matrix Multiplication: Performance}

Figure~\ref{fig:gb200emulationperformance} illustrates the performance characteristics of both native FP32 SGEMM (top) and BF16x9 emulated SGEMM (bottom).  The same color scale is used for both tables, to unify the two.  Given that peak BF16 MMA performance is 28$x$ greater than that of peak FP32 MMA performance on GB200, it is not surprising that the factor of 9$x$ overhead for BF16x9 emulation results in speed-ups as great as 3$x$ for some matrix sizes.  In these heatmaps, where we multiply $\textbf{A}_{[M\times K]}$ and $\textbf{B}_{[K\times N]}$, we set the M and N dimensions for matrix multiplication to be equal and vary the dimensions to cover a wide range of matrix sizes and aspect ratios.  The matrices were initialized with random data having a mean value of 0.0 and a standard deviation of 1.0.

While bandwidth-bound (low arithmetic intensity) regions show a performance disadvantage for BF16x9 SGEMM, the vast majority of the matrices tested saw improved efficiency with emulation.  Further, we note that libraries can easily be configured to utilize emulation only in cases where it will provide a performance benefit and that is the default behavior of this library.  However, to unambiguously illustrate performance differences, we utilized an environment variable to engage emulation in all cases for the lower chart.

\begin{figure}[h] \label{fig:gb200emulationperformance}
    \centering
    \includegraphics[width=1.0\linewidth]{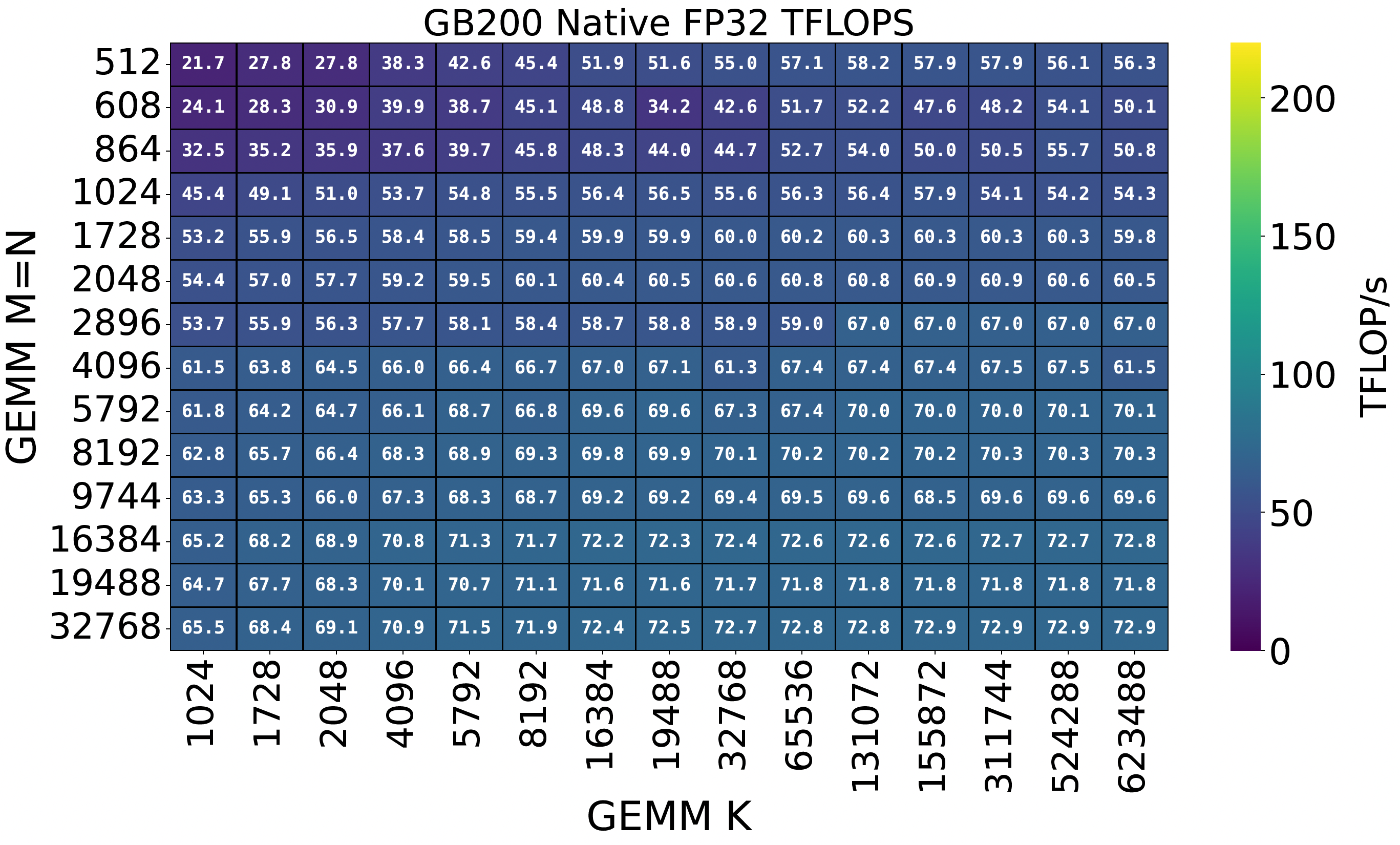}
    \includegraphics[width=1.0\linewidth]{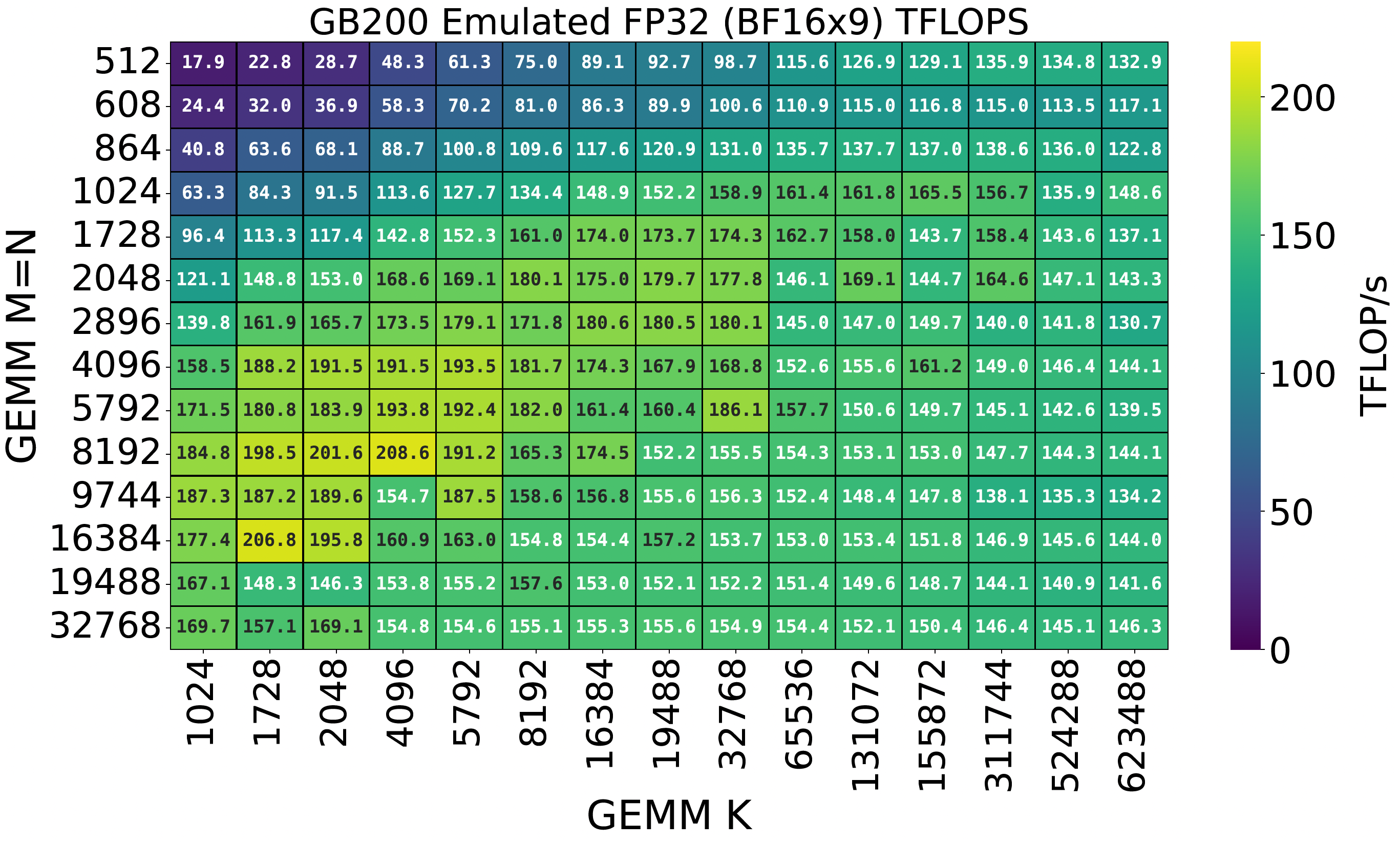}
    \caption{Comparison between native FP32 SGEMM (top) and BF16x9-emulated SGEMM (bottom) performance on GB200.  Performance is expressed in TFLOPS.  In some cases, BF16x9 performance is 3$x$ that of FP32.}
    \label{fig:gb200emulationperformance}
\end{figure}

\subsection{Matrix Multiplication: Power Efficiency}
\begin{figure}[h] \label{fig:b200emulationperformance}
    \centering
    \includegraphics[width=1.0\linewidth]{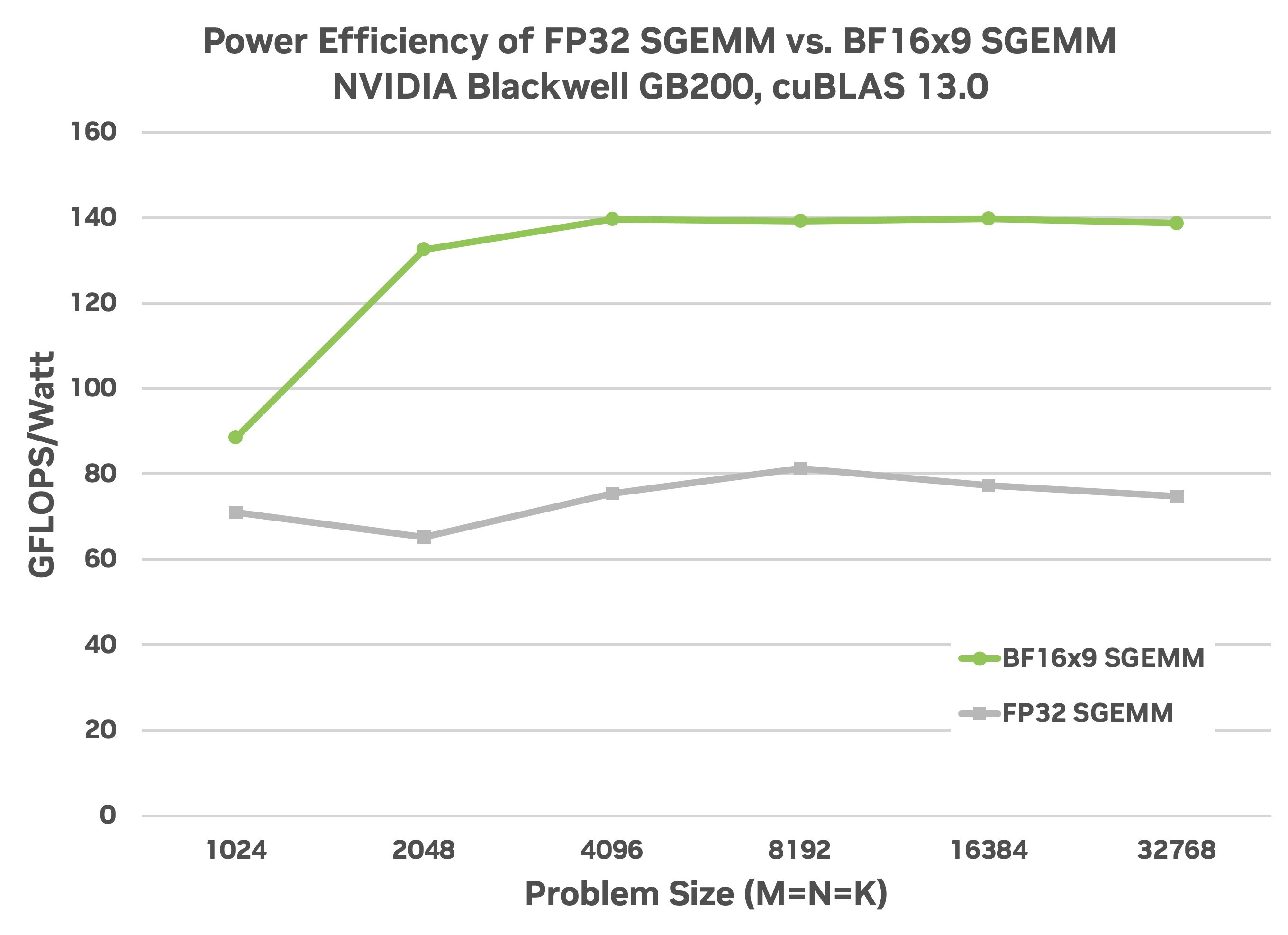}
    \caption{Power efficiency (GFLOPS/Watt) comparison between native (FP32) and emulated (BF16x9) SGEMM on GB200 (higher is better).}
    \label{fig:gb200emulationpower}
\end{figure}

In this section, we review the power requirements for native FP32 SGEMM vs. BF16x9 SGEMM.  In Figure~\ref{fig:gb200emulationpower}, we illustrate the 
measured GFLOPS/Watt on an NVIDIA Blackwell GB200. These results are based on running the NVIDIA System Management Interface (nvidia-smi) to monitor the power draw of these GPUs. We used cuBLAS 13.0, and filled our matrices with data produced via a normal distribution with a mean of 0.0, a standard deviation at 1.0, and enough iterations to provide reliable readings (smaller problems ran for more iterations). Runs were done for the GEMM case of $\alpha=1.0$, and $\beta=0.0$. In general, native FP32 SGEMM drew less instantaneous power, but, because it took more than twice the time to complete, the emulated BF16x9 SGEMMs of size 2048 or larger were approximately 40\% more power efficient, on average. 

For the application user, this power saving is practically free, since one does not have to change any APIs or set anything up, merely request BF16x9 emulation by setting an environment variable. The gains we see in power are significant and this was run with release 13.0, so the experiments should be repeatable for any user.

\subsection{Weather Simulation}

We ran spectral transforms on a GB200 cluster using 72 B200 GPUs connected through NVLink and detail the performance benefits of SGEMM emulation below. We performed a performance study of a ~1 km global resolution (TCo11999; 60 vertical levels) example and measured the total time of a forward and a backward transform. Using BF16x9 leads to a 2.4x speedup for the GEMM component, when compared to native FP32. This component consists of thousands of different GEMMs, of various sizes, executed inside of a large CUDA graph.  As the GEMMs are 50\% of the total runtime in the FP32 version, the overall speedup is 1.4$x$. After this improvement, the GEMMs only contribute to 31\% of the run time, distributing execution time evenly across the three main components.\\

\begin{figure}[h] \label{fig:weatherperf}
    \centering
    \includegraphics[width=\linewidth]{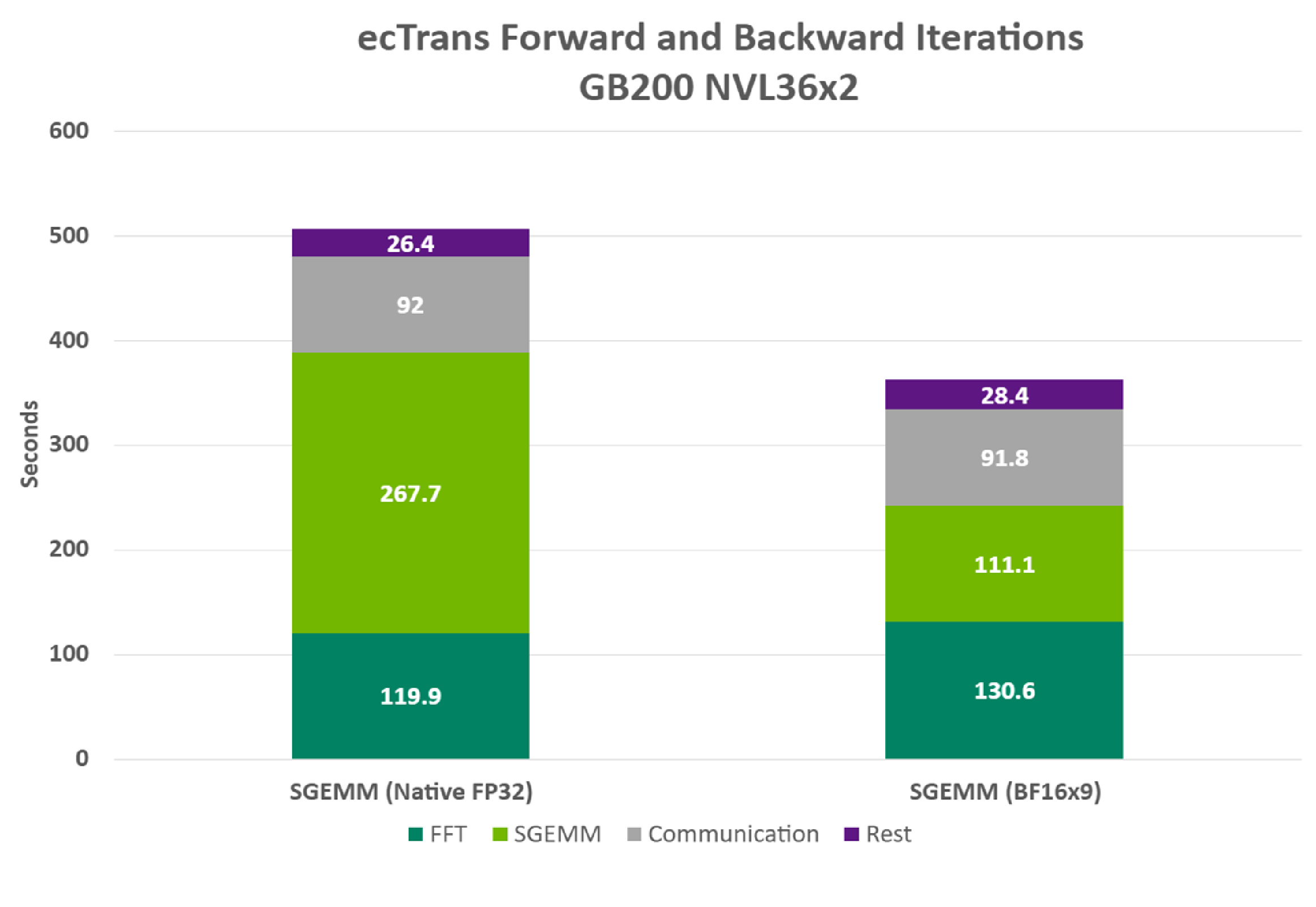}
    \caption{Comparison of the performance of ecTrans on an NVIDIA GB200 cluster NVL72, using FP32 and BF16x9 implementations of SGEMM. The SGEMM component speed-up ratio of BF16x9:FP32 is 2.4$x$, leading to an end-to-end speed-up of 1.4$x$.}
    \label{fig:weather1perf}
\end{figure}

\subsection{Quantum Chemistry}
\begin{figure}[htbp]
    \centering
    \includegraphics[width=\columnwidth]{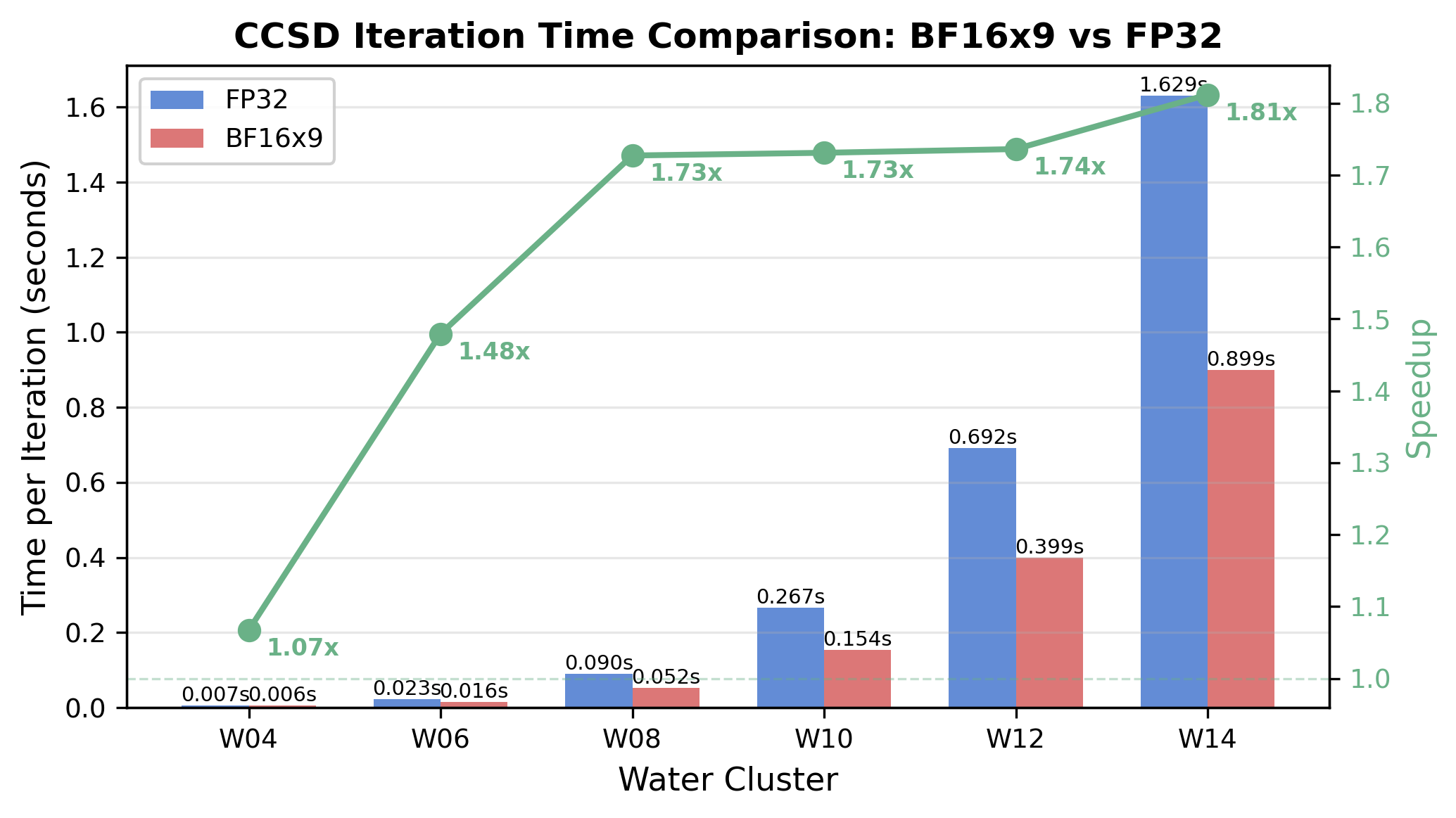}
    \caption{Comparison of CCSD iteration execution times for water clusters using native and emulated FP32.}
    \label{fig:ccsd_performance}
\end{figure}

We have measured the performance for the water-cluster benchmark described in Section~\ref{sec:ccsd-accuracy}. We find that a significant speedup is attained in CCSD iterations for all of our benchmarks. A CCSD iteration for the smallest cluster considered, water-4, shows a modest increase in performance by 7\% when using emulated FP32. However, this molecule is near the lower bound of molecule sizes that may be of interest. For most other clusters, a speedup factor of at least 1.73x is measured, with the CCSD iteration for water-14 benefiting from a 1.81x speedup. The iteration execution times and speedups that we have measured can be seen in Fig.~\ref{fig:ccsd_performance}. As molecule size increases, the CCSD iteration is dominated by the leading term of the form $A^{ab}_{ij} = \sum_{cd}t^{cd}_{ij} \sum_Q B_{ac}^Q B_{bd}^Q$. This term is calculated by exploiting symmetries as discussed in~\cite{DePrince_Sherrill_2013}. We find that this term asymptotically approaches a speedup of around 1.9x when measured in isolation using synthetic data, indicating that a full CCSD iteration should approach the same asymptote.

\section{Conclusions and Future Directions}\label{sec:concl}

The work presented in this paper demonstrates FP32 emulation, via BF16x9, working in a robust library setting.  This robustness extends to handling of both denormals and non-numbers (\texttt{NaN}s and \texttt{Inf}s) as they are treated with FP32 SGEMM.  Further, due to the capabilities of NVIDIA's Blackwell GPUs, including high throughput and integrated scaling capabilities, the performance of emulated SGEMM has been shown to greatly exceed that of native (FP32-based) SGEMM, without sacrificing accuracy.  To that end, we have provided evidence of accuracy with both synthetic test cases with prescribed numerical properties and with real-world application results.  Additionally, we have shown that the library code, fully API-compatible with the standard SGEMM, can yield a factor of up to 3.0$x$ performance improvement, with a simultaneous 40\% power savings, when compared to FP32-based SGEMM.

We have a number of potential extensions that we are exploring.  To reach even greater levels of performance, we are investigating the coding and API changes required to furnish users who can tolerate potentially less accurate modes of emulation with the ability to do so in a non-intrusive fashion.  We are also considering how to extend this work both up and down both the precision and accuracy range, as is done with the parameterized Ozaki scheme~\cite{Uchino_2025} used for the emulation of FP64 matrix multiplication (DGEMM).  Another approach in this domain that we are considering is the use of BF16 compute to emulate FP64 when the exponent range of the FP64 values is restricted to a domain that lends itself to this approach, as the challenge with this methodology has to do with the extended exponent range of FP64.  Finally, in order to provide an even greater boost to applications, we are researching the ways in which we can lower the crossover point (matrix dimensions) at which BF16x9 SGEMM outperforms native FP32 SGEMM.


\clearpage
\bibliographystyle{acm}
\bibliography{ourbib}
\vspace{12pt}

\end{document}